\documentclass[12pt, eqsecnum]{revtex4}
\usepackage{graphicx}
\begin{document}
\def\gtap{\mathrel{ \rlap{\raise 0.511ex \hbox{$>$}}{\lower 0.511ex
   \hbox{$\sim$}}}} \def\ltap{\mathrel{ \rlap{\raise 0.511ex
   \hbox{$<$}}{\lower 0.511ex \hbox{$\sim$}}}}
\newcommand{\beq}{\begin{equation}}
\newcommand{\dd}{\partial}
\newcommand{\eeq}{\end{equation}}
\newcommand{\bea}{\begin{eqnarray}}
\newcommand{\eea}{\end{eqnarray}}
\newcommand{\lsim}{\stackrel{<}{\scriptstyle \sim}}
\newcommand{\gsim}{\stackrel{>}{\scriptstyle \sim}}
\newcommand{\La}{{\cal L} }
\newcommand{\half}{\frac{1}{2} }
\newcommand{\eq}[1]{eq.(\ref{#1})}
\newcommand{\dpar}[2]{\frac{\partial #1}{\partial #2}}
\newcommand{\vpar}[2]{\frac{\delta #1}{\delta #2}}
\newcommand{\ddpar}[2]{\frac{\partial^2 #1}{\partial #2^2}}
\newcommand{\vvpar}[2]{\frac{\delta^2 #1}{\delta #2^2}}
\newcommand{\eV}{\mbox{$ \ \mathrm{eV}$}}
\newcommand{\KeV}{\mbox{$ \ \mathrm{KeV}$}}
\newcommand{\MeV}{\mbox{$ \ \mathrm{MeV}$}}
\newcommand{\probm}{\mbox{$ \ \langle P_m \rangle$}}
\newcommand{\bz}{\bar{z}}
\newcommand{\sg}{\sigma}
\newcommand{\tw}{\tilde{\omega}}
\newcommand{\rla}{\longrightarrow}


\title{ Complex Classical Mechanics of a QES Potential }
\date{\today}

\author{Bhabani Prasad Mandal}
\affiliation{ Department of Physics, Banaras Hindu University, Varanasi-221005, India}
\email{bhabani.mandal@gmail.com}

\author{Sushant S. Mahajan}
\affiliation{ Department of Physics, Indian Institute of Technology, Banaras Hindu University, Varanasi-221005,
 India}
\email{sushant.mahajan.itbhu@gmail.com}


\begin{abstract}
We consider a Parity-time (PT) invariant non-Hermitian quasi-exactly solvable (QES) potential which exhibits PT phase transition.
 We numerically study this potential in a complex plane classically to demonstrate different quantum effects. The particle with real energy
makes closed orbits around one of the periodic wells of the complex potential depending on the initial condition. However interestingly
the particle can have
open orbits even with real energy if it is initially placed in certain region between the two wells on the same side of the
imaginary axis. On the other hand when the particle energy is complex the trajectory is open and the particle tunnels back
and forth between two wells which are separated by a classically forbidden path. The tunneling  time is calculated for different  pair of
wells and is shown to vary inversely with the imaginary component of energy.  At the classical level unlike the analogous quantum situation we do not see any qualitative differences in the features of the particle dynamics for  PT symmetry broken and unbroken phases.
\end{abstract}

\maketitle
\section{Introduction}
Consistent quantum theory with unitary time evolution and probabilistic interpretation for
certain classes  of non-Hermitian systems have been the subject of intrinsic research
in frontier physics over the last one and half decade \cite{ben,rev,rev1}. The huge success of complex
quantum theory \cite{pts,km,pts1,pts2,pts3,pts4,pts5,pts6,pts7,pts8,pts9,pt91,pt92,pt93,pt94, pts10} has lead to its extension to many other branches of
physics \cite{qft,qft1,qft2,opt,pte,pha1,baz,de,pha2}. In particular, its
application to quantum optics is the most noticable, where break down of PT-symmetry has
been observed experimentally \cite{opt,pte}. More recently, this formulation has been extended to classical
systems \cite{cm,af1,cm1,cm2,cm3}. Quantum mechanics and classical mechanics are two completely different theories and
provide profoundly dissimilar description of physical systems. However, Bohr's correspondence
principle states that quantum particles behave like classical ones when the quantum number
is very high. Keeping this in mind, classical systems have been investigated in a complex
plane. Correspondence between quantum and classical systems
becomes more pronounced in the complex domain \cite{cm,af1}. Remarkably, a particle with
complex energy exhibits tunneling like behavior which is usually realized in the quantum
domain. This tunneling behavior of a classical particle with complex energy is well
demonstrated in Ref. \cite{cm,af1}.  They have shown that a classical particle can tunnel
from one classically allowed region to another allowed region separated by a classically
forbidden path. Several other works in this field are devoted to study the nature of the
trajectory of a classical particle with complex energy \cite{cm1,af1,cm2,cm3}. Attempts have been made to study the
effect of spontaneous PT-symmetry breaking in the particle trajectories of analogous complex
classical models. However the study of complex classical system is only restricted to a few models.

The purpose of the present article is to extend these works further to investigate the different
behaviors of a complex classical system. We have chosen a complex classical system corresponding
to  a QES \cite{qes,qes1,qes2,qes3}, non-Hermitian PT-symmetric system \cite{km} described by the potential
$V(x)=-(\zeta\cosh 2x-iM)^2$.
We numerically study the dynamics of a classical particle moving in the complex xy-plane subjected to this potential which consists of periodic wells situated
to the left and right side of the imaginary axis
to demonstrate the quantum tunneling effect and the trajectory of the particle in different situations. For real energy, the particle
makes closed orbits around one of the wells depending on the initial condition. However, surprisingly the particle can have
open orbits even with real energy if it is initially placed in a certain region between the two wells on the same side of the
imaginary axis. On the other hand when the particle energy is complex, the trajectory is open and the particle tunnels back
and forth between two wells which are separated by a classically forbidden path. In all the situations the particle trajectory never crosses itself.
At the classical level, unlike the analogous quantum situation, we do not see any qualitative differences in the features of the particle dynamics for $M$-even ( PT symmetry broken phase) and $M$-odd ( PT symmetry unbroken phase).  The tunneling  time is calculated for different  pairs of
wells and is shown to vary inversely with the imaginary component of energy, similar to a situation which is usually valid in the quantum domain.

Now, we state the plan of the paper. The QES potential  and its solutions for broken and unbroken PT symmetry cases are reviewed in Sec.II. We discuss the
classical mechanics of this potential in a complex plane and obtain the trajectories of the classical particle in different situations in Sec.III. Variation of
tunneling time with imaginary part of the energy of the particle is discussed in sec.IV, and section V is kept for conclusions.

\section{The QES system}
The complex QES system is described by the Hamiltonian
\begin{equation}
H= p^2-(\zeta\cosh 2x -iM)^2
\end{equation}
where $\zeta$ is an arbitrary real parameter and $M$ is an integer and we have considered
$2m=\hbar=1$.
This Hamiltonian is  symmetric under combined parity and time reversal transformation.
Parity transformation in this particular system is taken in the general form as
$x\rightarrow a -x $, where $ a=i\frac{\pi}{2}$.
This system is shown to be a QES system. The first $M$ energy levels along with the corresponding
eigenfunctions are calculated exactly for any specific integer value of $M$. Further more,
this QES system shows another remarkable property relevant to PT symmetric non-Hermitian
system, i.e. for even values of $M$, all the eigenvalues are complex for any arbitrary value of $\zeta$. On the other hand, for odd $M$, all the eigenvalues are real if $\zeta \leq \zeta_c$.
In other words, system is always in broken PT phase for even values of $M$, and shows PT phase
transition when $M$ is odd.
The energy eigenvalues for this system is calculated using Bender-Dunne (BD) polynomial methods \cite{bd}. The zeros of BD polynomials give the energy eigenvalues for the QES system.
Some of the low lying levels are listed as follows.
\begin{eqnarray}
M =2&&\\ \nonumber
&&E_{\pm} =3-\zeta^2\pm 2i\zeta;\ \  \phi_+ = A\cosh x \ \ \& \ \ \phi_-= B\sinh x \\ \nonumber
M= 4 &&\\ \nonumber
E_{\pm}^1 &=& 11-\zeta^2 -2i\zeta \pm\sqrt{1-i\zeta-\zeta^2};\ \ \phi_{\pm}^1= A\sinh 3x +B\sinh x
\\ \nonumber
\mbox{with} && \frac{A}{B} = \frac{E-7+\zeta^2}{2i\zeta}
\end{eqnarray}

We can  have PT phase transition when, $M$ is odd
\begin{eqnarray}
M=1 &&\\ \nonumber
&&E_{\pm} = 1-\zeta^2;\ \  \phi = A(\mbox{constant}) \\ \nonumber
M=3&&\\ \nonumber
&&E_{\pm} = 7-\zeta^2  \pm\sqrt{1-4\zeta^2};\ \ \ \mbox{with} \phi_{\pm}= A\cosh 2x +iB\\ \nonumber
&& E= 5-\zeta^2, \ \ \mbox{with } \ \phi = C\sinh 2x
\\ \nonumber
\mbox{with}&& \frac{A}{B} = \frac{4\zeta}{E-9+\zeta^2}
\end{eqnarray}
Note that for odd $M$, energy eigenvalues are real subject to the condition $\zeta\leq\zeta_c$ and the wave functions are
also eigenstates of the $PT$ operator. PT phase transition occurs at $\zeta=\zeta_c$. This conclusion continues to be valid for higher
odd values of $M$\cite{km}.

\section{ Classical Mechanics in the Complex Plane}

\begin{figure}[t]
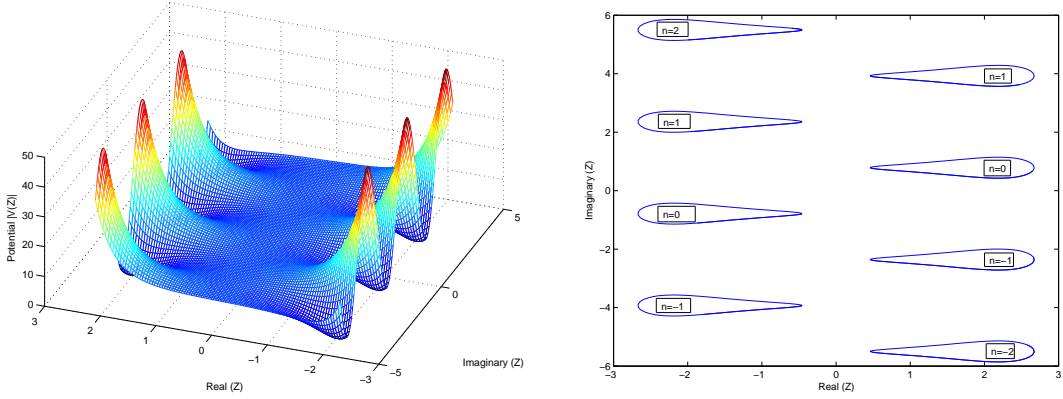

\includegraphics[width=7cm]{Meshc.epsc}
\includegraphics[width=7cm]{RealE.epsc}
\caption{\label{fig1}{\bf Left:} The magnitude of potential in the complex plane for $\zeta=0.1$ and $M=3$. The potential wells corresponding to the real energy orbits  are distributed periodically, centered at $x=2.04750, y=\frac{4n+1}{4}\pi$ on the right, and $x=-2.04750, y=\frac{4n-1}{4}\pi$ on the left of the imaginary axis. {\bf Right:} The closed orbits traced by the particle when placed at different places in the complex potential with Real Energy $E=0.8$. This Fig. also indicates the positions of the wells.}
\end{figure}

\begin{figure}
\includegraphics[width=8cm]{potM.epsc}
\caption{\label{fig2}The double well potential for real x and its variation with the parameter M for $\zeta=0.1$}
\end{figure}

A classical particle with real energy $E$ is not allowed to travel in the region
where the potential energy $V(x)> E$. However, this restriction is relaxed when we consider the particle in a complex plane with complex energy $E_1+iE_2$ as
\begin{equation}
E_1= (p_1^2-p_2^2) +V_1;\ \ \  E_2=2p_1p_2 +V_2
\label{e1e2}
\end{equation}
where complex momentum $p=p_1+ip_2$ and we write the potential $V=V_1+iV_2$. Now since $p_1$ and
$p_2$ can have any value from $-\infty$ to $\infty$ there is no restriction as such on the particle to be bound in a particular region of space. Particle is allowed to move anywhere in the complex plane as long as the Eq. (\ref{e1e2}) is satisfied. This is the prime reason for a classical particle with complex energy to travel through classically forbidden regions.
However, even though the particle can exist anywhere in the complex plane, it prefers
the region with lower energy. The particle follows a definite trajectory depending on initial
conditions. Open trajectories with local random walk type motion have been shown for a classical particle with complex energy \cite{cm}.
Depending on the value of the complex energy, classical particle can be delocalized and move freely in the potential.
 However, a classical particle with real energy generally moves  in a closed orbit in the complex plane \cite{cm2}.

 \begin{figure}
\includegraphics[width=6cm]{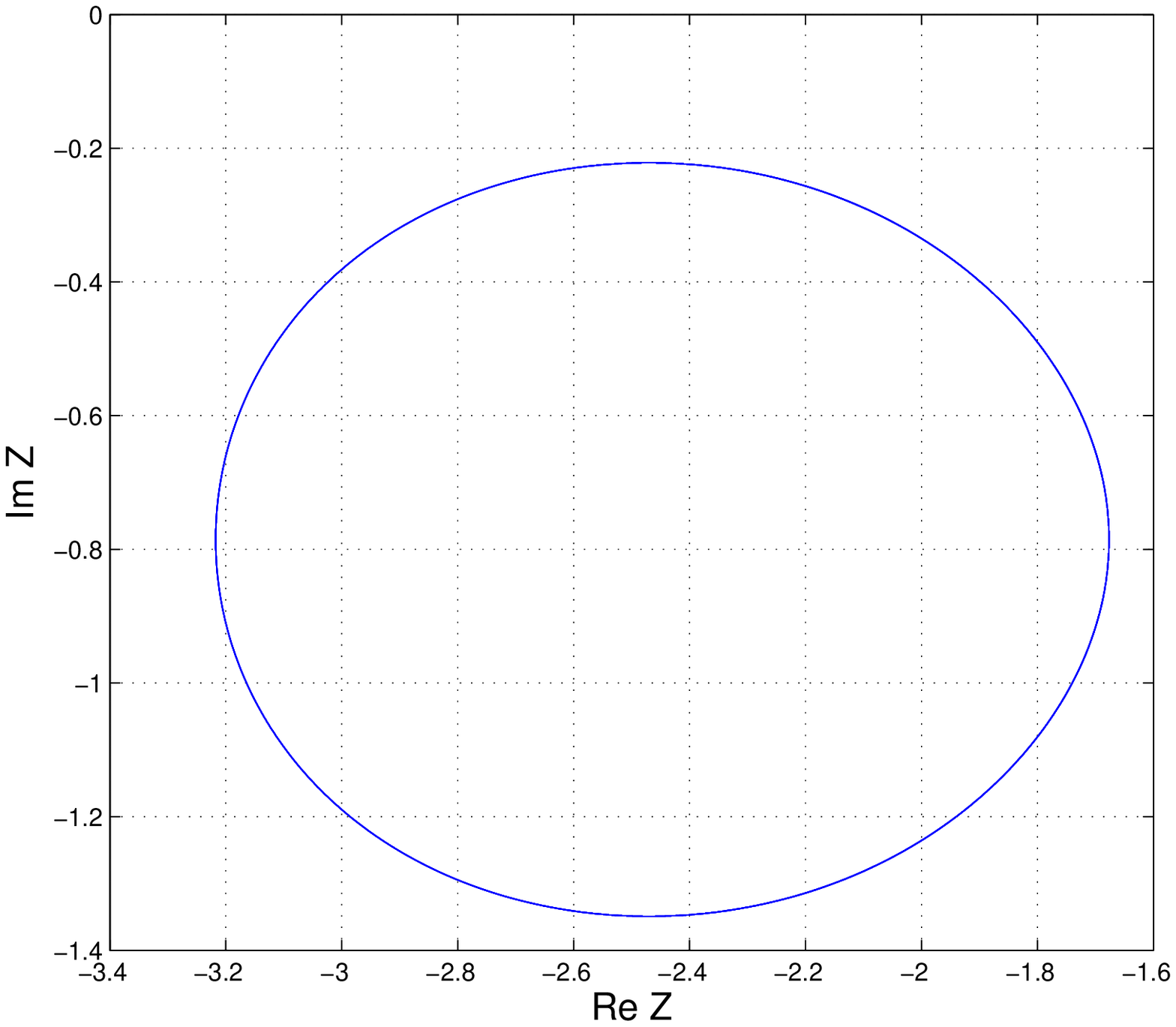} \ \ \ \ \
\includegraphics[width=6cm]{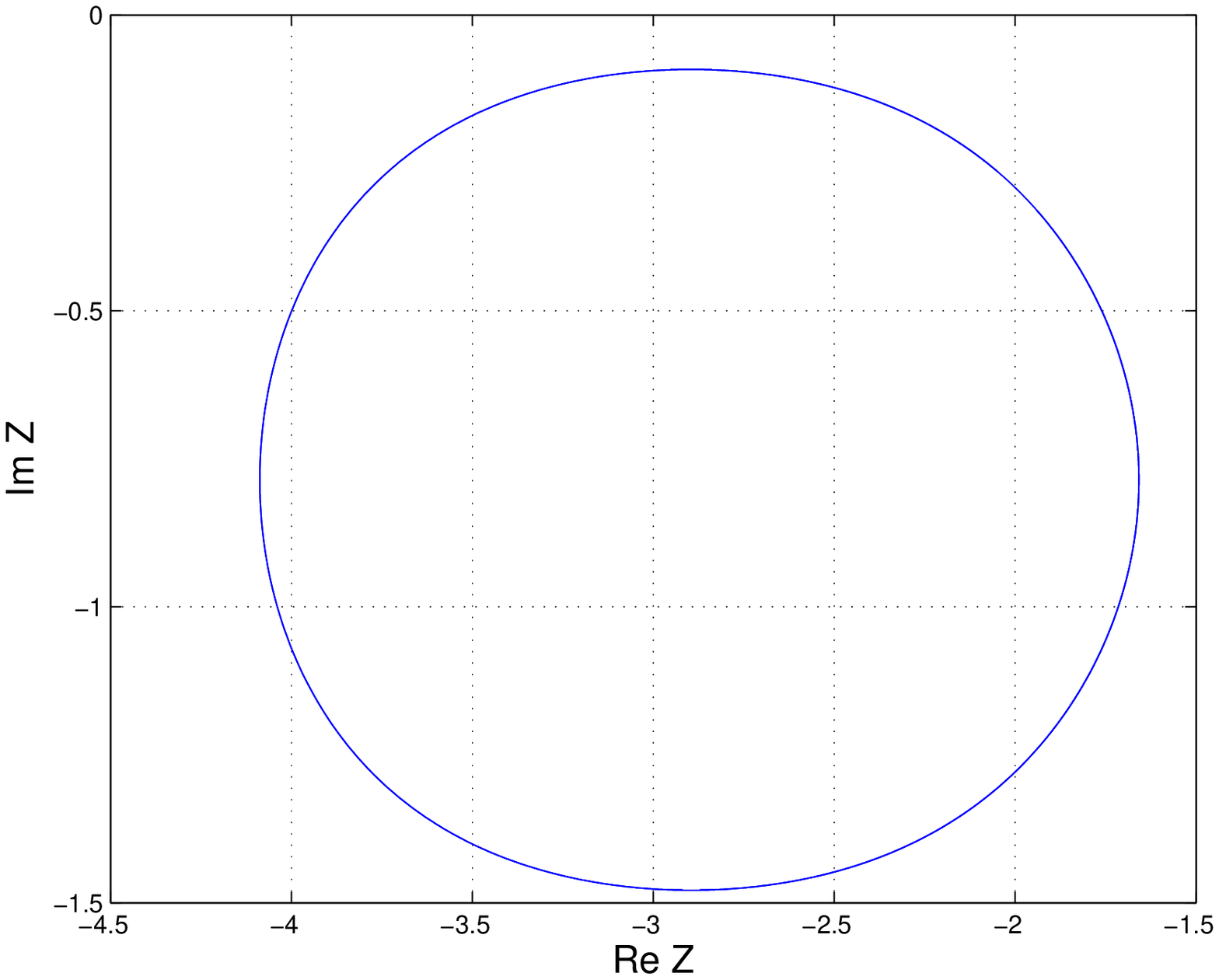}
\caption{For $\zeta=0.1$, $M=3$, $E=0.8$. Trajectories of the particle when initial position is $Re(Z)=-2.04750$ and $Im(Z)=y$.
{\bf Left:} $y=n_{0}+0.4740$; \ \ \ \ \ {\bf Right:} $y=n_{0}+0.52$}
\includegraphics[width=6cm]{y+0.54} \ \ \ \ \ \ \ \
\includegraphics[width=6cm]{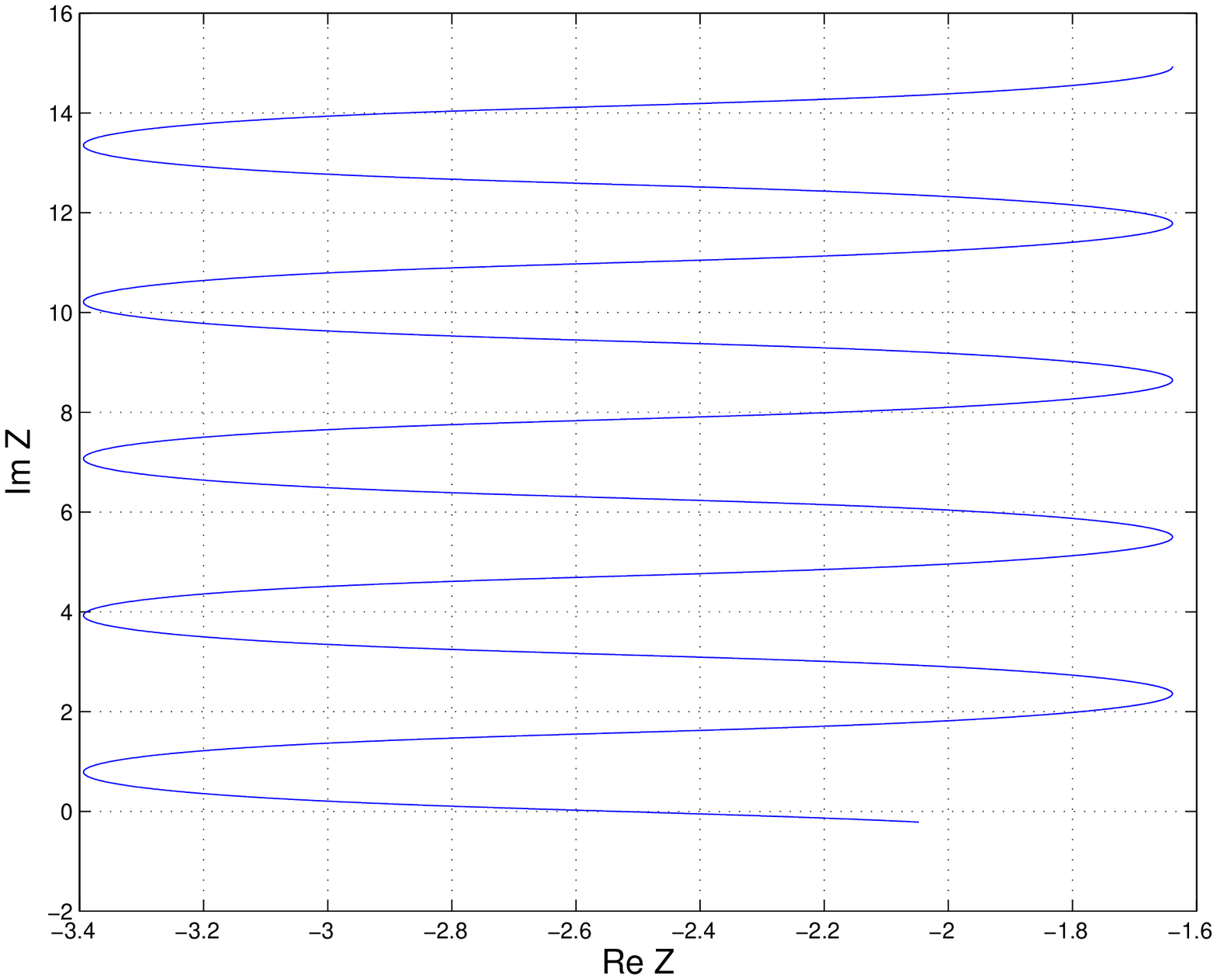}
\caption{For $\zeta=0.1$, $M=3$, $E=0.8$. Trajectories of the particle when initial position is $Re(Z)=-2.04750$ and $Im(Z)=y$. {\bf Left:} $y=n_{0}+0.54$; \ \ \ \ \ {\bf Right:} $y=n_{0}+0.57$}
\end{figure}

\begin{figure}
\includegraphics[width=6cm]{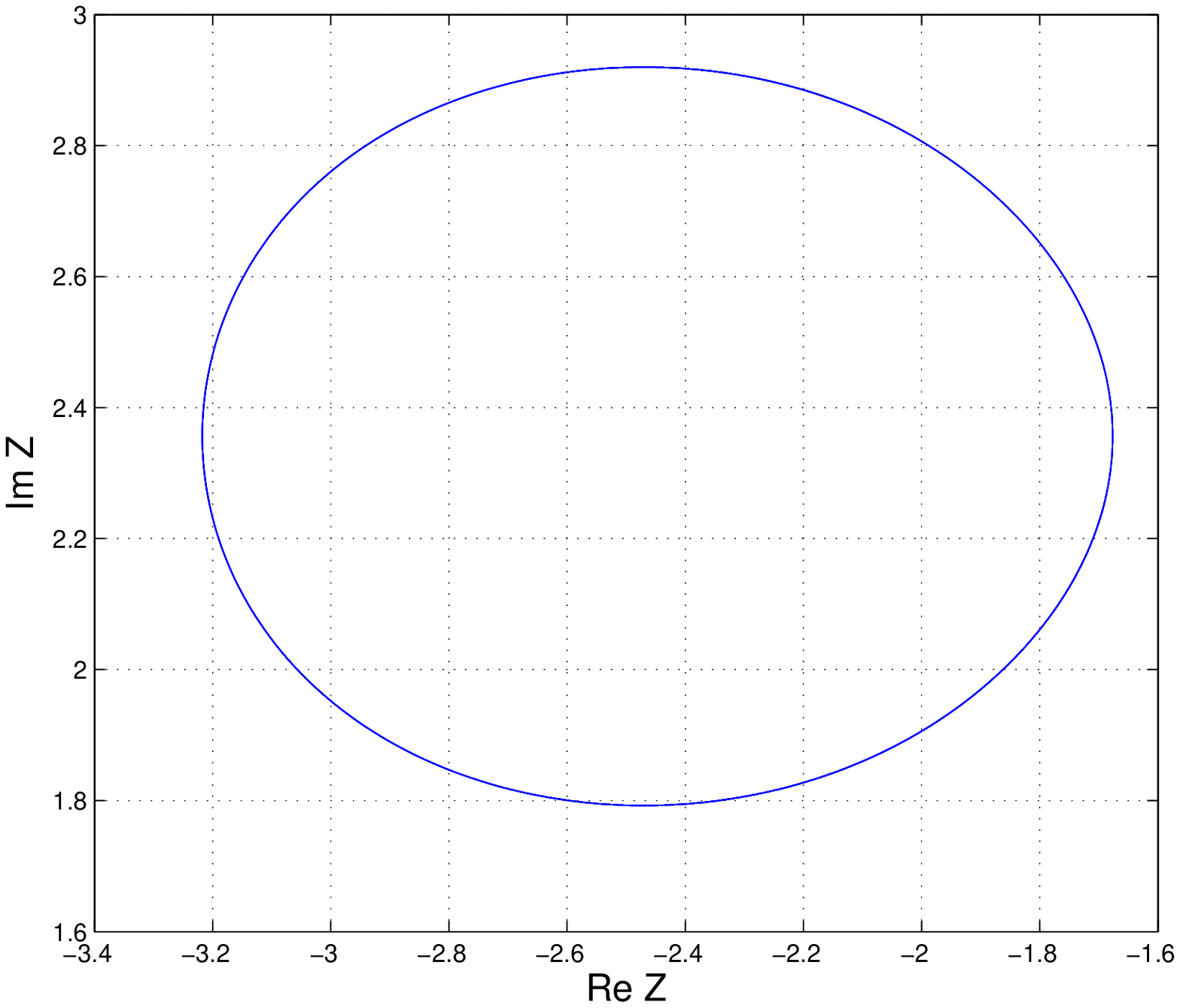}
\includegraphics[width=6cm]{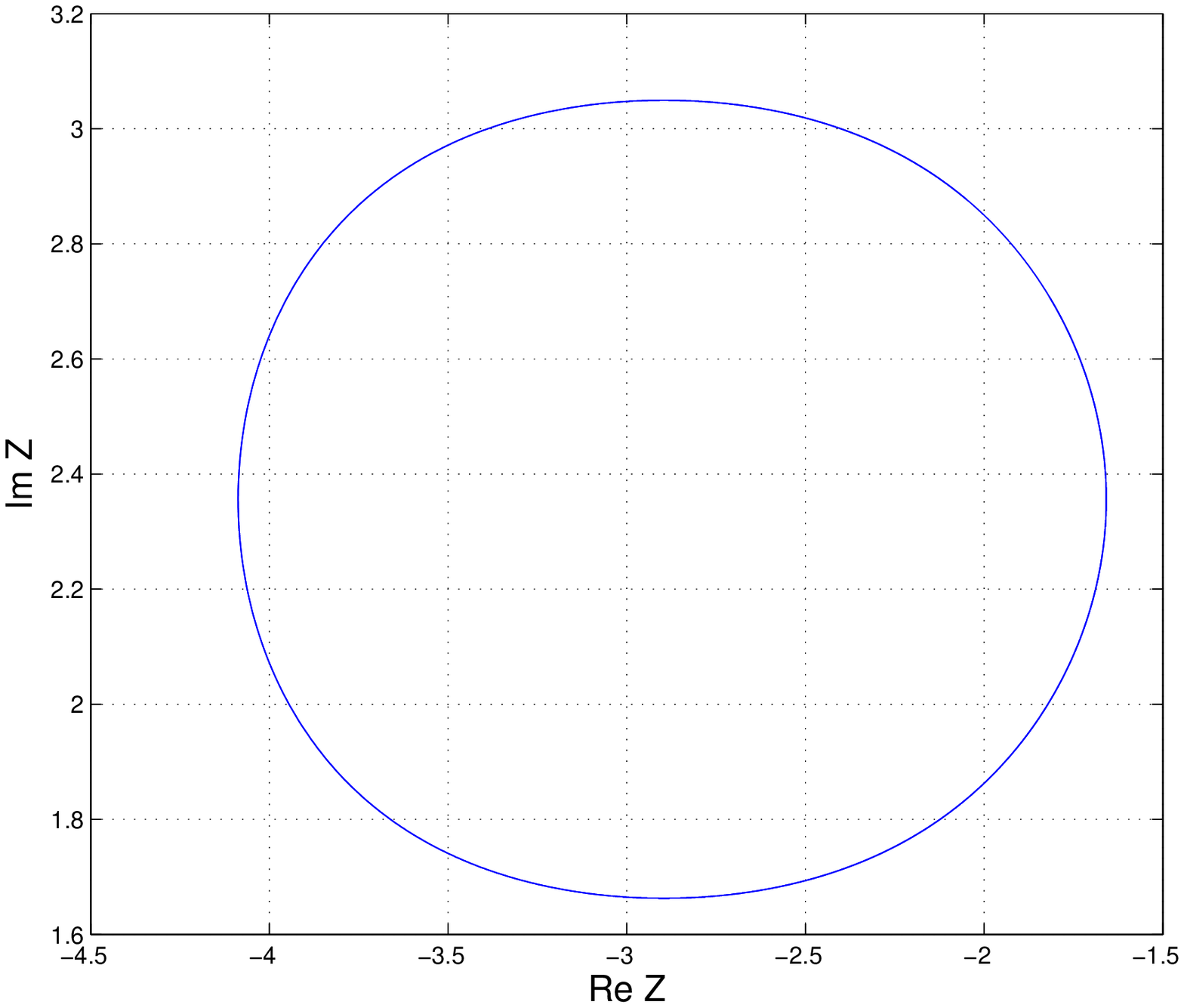}
\caption{For $\zeta=0.1$, $M=3$, $E=0.8$. Trajectories of the particle when initial position is $Re(Z)=-2.04750$ and $Im(Z)=y$. {\bf Left:} $y=n_{0}+\pi+0.4740$; \ \ \ \ \ {\bf Right:} $y=n_{0}+\pi+0.52$}
\includegraphics[width=6cm]{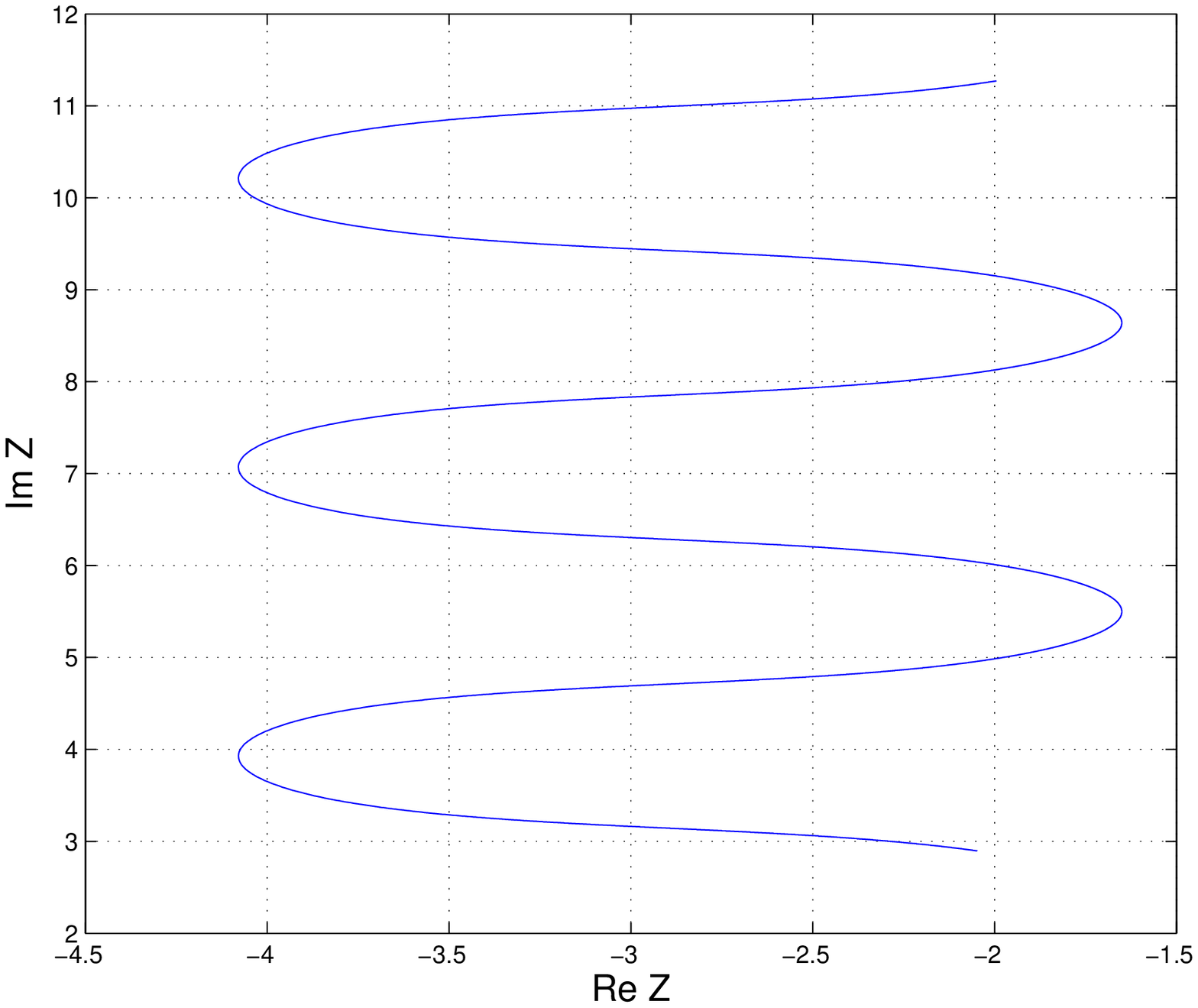}
\includegraphics[width=6cm]{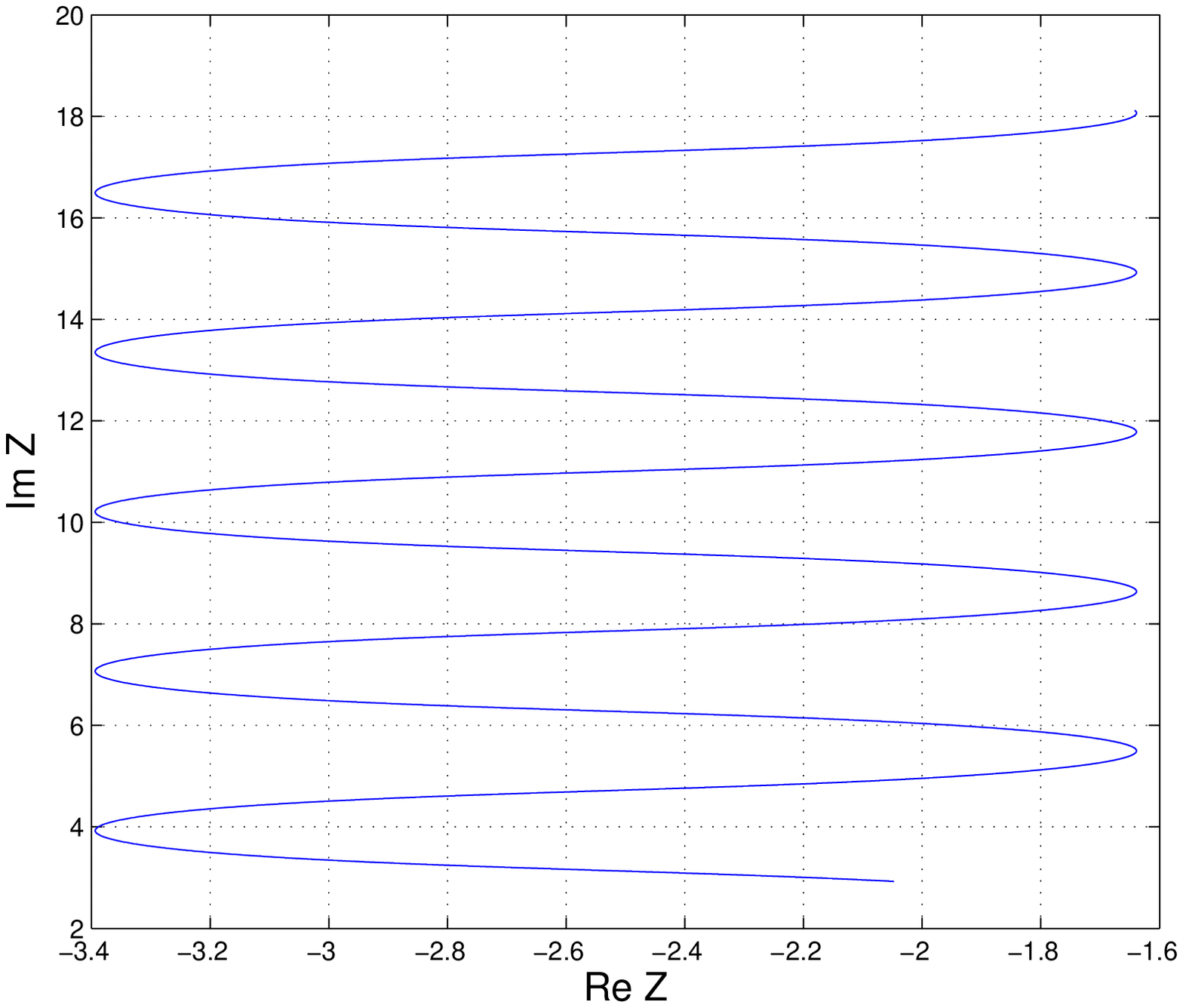}
\caption{For $\zeta=0.1$, $M=3$, $E=0.8$. Trajectories of the particle when initial position is $Re(Z)=-2.04750$ and $Im(Z)=y$. {\bf Left:} $y=n_{0}+\pi+0.54$; \ \ \ \ \ {\bf Right:} $y=n_{0}+\pi+0.57$}
\end{figure}

We consider the classical system of a particle moving under the influence of the above QES potential in a complex plane. The complexified potential  $V(z) = -(\zeta\cosh 2z -iM)^2 $ in the complex
plane $z=x+iy $ is shown in the Fig.1. The potential consists of a series of wells on both left and right
side of the imaginary axis [i.e. y-axis]. On the left of the imaginary axis, the wells are positioned at $y=\frac{4n-1}{4}\pi$ and on the right of imaginary axis
at  $y=\frac{4n+1}{4}\pi$, where n is an integer, as well as a label for the wells.
The  Hermitian counterpart of this potential
$V(x) = -(\zeta\cosh 2x -M)^2 $ for different values of $M$  with fixed value of $\zeta (=0.1)$ is plotted in Fig.2.
The central barrier height increases with $M$ values. We consider a particle in this complex QES potential. The dynamics of this particle are governed by the Hamilton's equation as
\begin{equation}
\dot{z}=\frac{\partial H}{\partial p}; \ \ \ -\dot{p}=\frac{\partial H}{\partial z}
\label{heq}
\end{equation}
where $H=\frac{p^2}{2m}+ V(z) $. We find the trajectories in the complex plane
 for different complex energies by solving the Eq. [\ref{heq}] for the above QES potential numerically. We solve this
 system taking the units such that, $2m=\hbar=1$. Since the imaginary part of energy is thought to arise through quantum mechanical uncertainty, this choice of units will give us the advantage of measuring the energy in multiples of $\hbar$ i.e. unit energy$= 1.0545 X 10^{-34} Joules$. To observe quantum effects, we have to go well beyond the visibility of the naked eye, and hence we have chosen a length scale of one nanometer. Choosing this system of units gives the units of mass and time to be milligram and second respectively. Thus, our test particle has a mass of half code units, i.e. 0.5 milligrams. This particle can safely be considered as a classical particle. All following discussions and figures in this paper display the quantities in these units.

 We see
 that particle orbits around the position of the different wells if the energy of the particle is real and the particle is placed sufficiently close
to the well [Figs. 3 \& 5]. However, the particle can have open orbits even with real energy if not placed sufficiently close to a well [Figs 4 \& 6].


\begin{figure}[h]
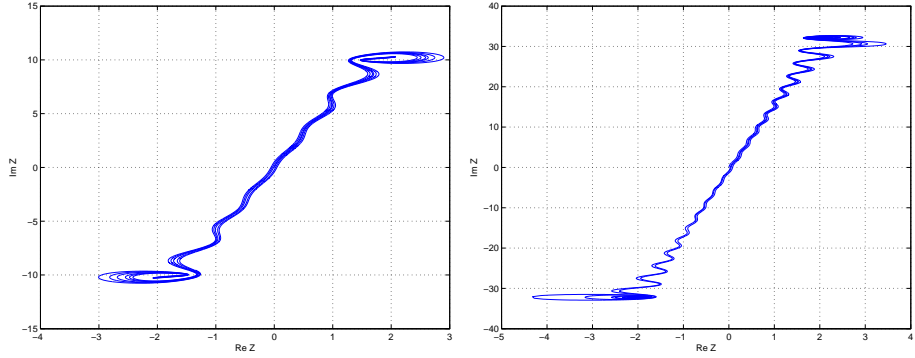

\includegraphics[width=6cm]{Z=0.1,M=2.epsc}
\includegraphics[width=6cm]{Z=0.1,M=3.epsc}
\caption{The trajectory of a particle with energy $ E=1+i$ in potential for $ \zeta=0.1$. {\bf Left }: when $ M=2$, the particle oscillates between the wells corresponding to $ n=+3$ on the right  $ n=-3$ on the left  of the imaginary axis.
{\bf Right}: when $ M=3 $, the particle oscillates between the wells corresponding to  $n=+10$ on the right and $n=-10$ on the left of the imaginary axis. }
\end{figure}

\begin{figure}[h]
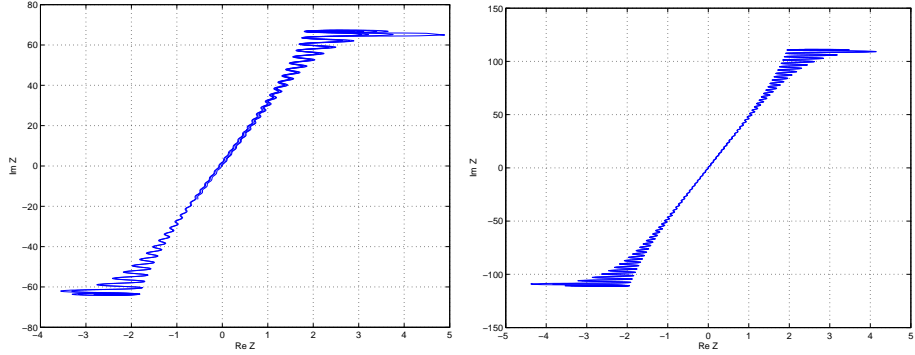

\includegraphics[width=6cm]{Z=0.1,M=4.epsc}
\includegraphics[width=6cm]{Z=0.1,M=5.epsc}
\caption{The trajectory of a particle with energy $E=1+i$ in potential for $\zeta=0.1$. {\bf Left:} when ${M=4}$, the particle oscillates between the wells corresponding to  $n=+21$ on the right and $n=-21$ on the left of the imaginary axis.
{\bf  Right:} when ${M=5}$, the particle oscillates between the wells corresponding to  $n=+35$ on the right and $n=-35$ on the left of the imaginary axis.}
\end{figure}

If we vary the distance of the starting point of the particle from the well along the imaginary axis (y-axis), the closed orbit opens up after a certain value.
For the first well ($n=0$ ) on the left of imaginary axis if particle is placed at $y < (n_0+.52988875)$ or $y > (n_0-.52988875)$ we have closed orbit of the particle. Where $n_0= -
\frac{\pi}{4}$, the position of first well along $y$ direction. However if the
particle is not placed sufficiently close $y\ge (n_0+.52988875)$ or $y\le (n_0-.52988875)$ to the first well, the particle will have open orbit even with real energy and will escape to
infinity without tunneling to any other well. We have illustrated this for the first and second well, i.e for   $n=0$ and $1 $ on the left side of the imaginary axis.
The second well is positioned at $y=\frac{3\pi}{4}\equiv n_1$
along imaginary axis. The particle escapes to infinity if it is placed beyond $y= n_1\pm.52988875$.




The most interesting results that we have obtained for complex energy of the particle are that:

1. It tunnels back and forth between the wells to the left and right of the imaginary axis.

2. When placed inside a well, the particle spirals out until it enters the classically forbidden region outside the well, then depending on the direction of its velocity, it spirals into the first well (on the other side of the imaginary axis) that it encounters in its path. The separation of the wells between which the tunneling takes place increases with $M$ values and also depends on the complex energy and on the parameter $\xi$.
This spiraling nature was also shown in Refs. \cite{cm1,cm2}

\begin{figure}[h]
 \includegraphics[width=6 cm]{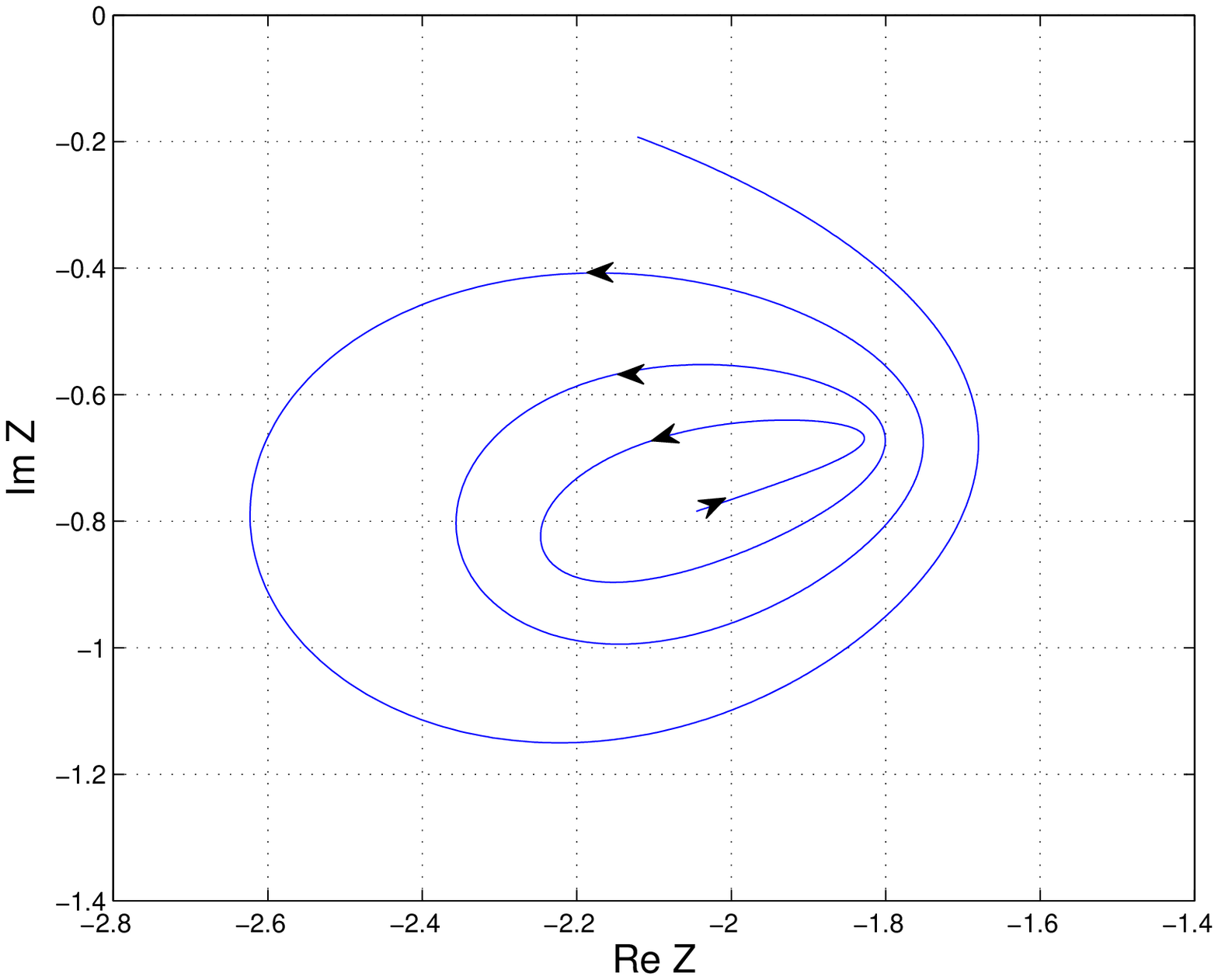}
 \includegraphics[width=6 cm]{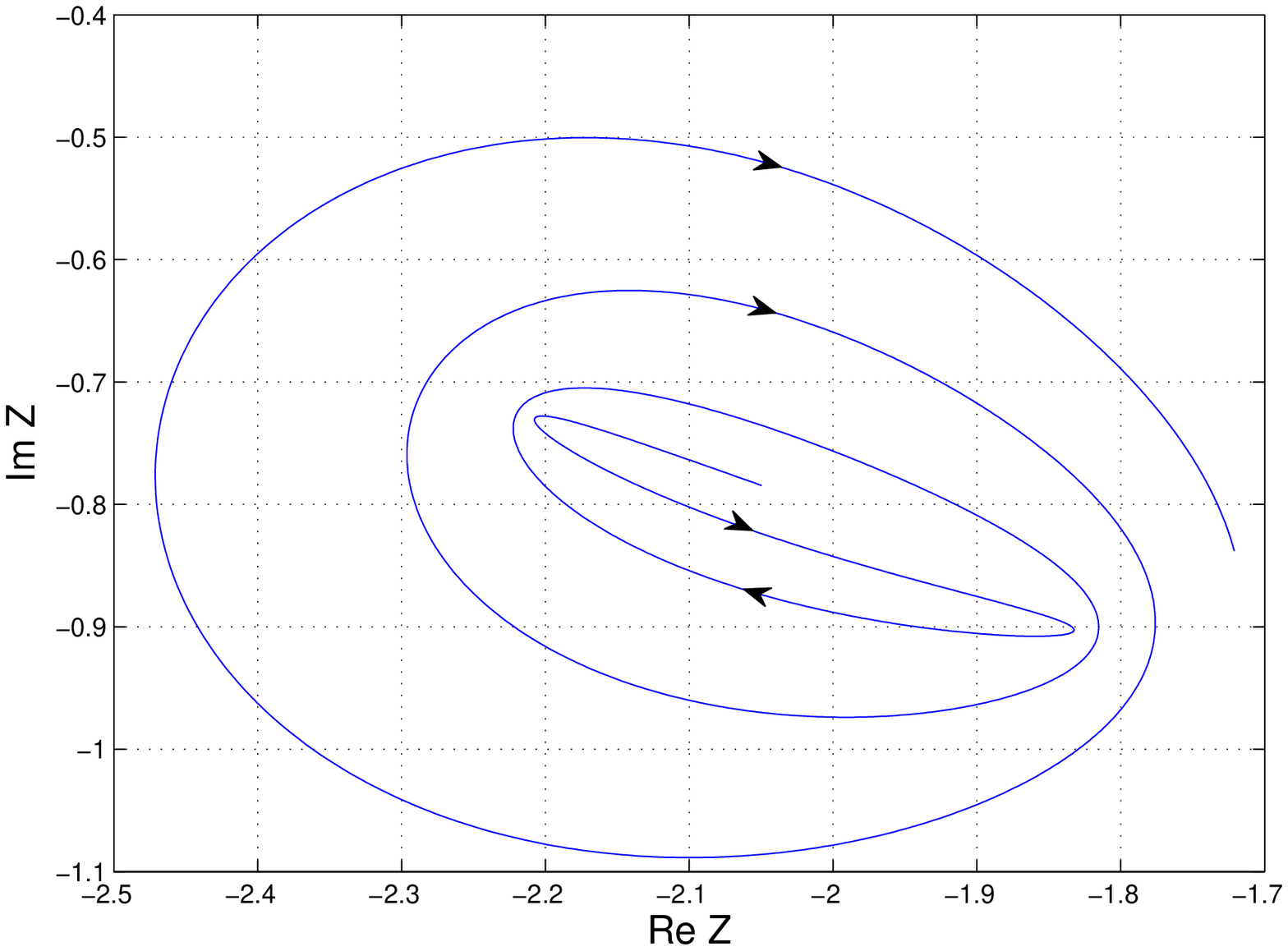}
 \caption{The particle spiralling outwards from the well at n=0 on the left side of the imaginary axis. {\bf Left}: for $E_2>0$ {\bf Right}: for $E_2<0$ }
 \label{direction}
\end{figure}

3. When the particle is spiraling in, towards the center of a well, in all cases with $E_2>0$, we observe that its motion is clockwise, and, anticlockwise when it spirals out from the well.
Whereas for $E_2<0$, the spiral motion is anticlockwise inwards and clockwise
outwards. The outward spiralling of the particle from the well at n=0 on the left
side of the imaginary axis is shown in figure \ref{direction}. For $E_2<0$, the right side well has $n<0$ and the left side well has $n>0$.

4. For the results in figs 7-10, we tried 3 different initial conditions. (i) Placing the particle in the well on the left side,
(ii) placing the particle in the well on the right side, and (iii) placing the particle at the origin. All the three initial conditions give qualitatively the same result.
The only difference is that if the particle is placed at the origin, and it is found to tunnel between two wells
n=-a on the left and n=+a on the right, then placing the particle initially in the well corresponding to n=-b on the left
will make the particle tunnel between the wells n=-b on the left and n=-b+2a on the right. Similarly, if the particle
is initially placing in the well n=+b on the right, then it will tunnel between n=+b on the right and n=+b-2a on the left.
For simplicity, we have used origin as the starting point.

5. We have also observed that the trajectory of the particle in this complex potential never crosses itself.

\begin{figure}
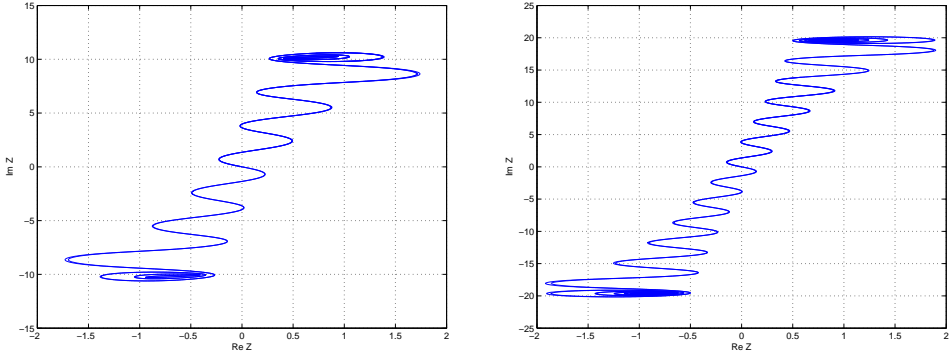

\includegraphics[width=6cm]{Z=1,M=2.epsc} \ \ \ \
\includegraphics[width=6cm]{Z=1,M=3.epsc}
\caption{The trajectory of a particle with energy $E=1+i$ in potential for $\zeta=1$.
{\bf Left:} when $M=2$, the particle oscillates between the wells corresponding to $n=+3$ on the right and $n=-3$ on the left of the imaginary axis.
 {\bf Right:} when $M=3$, the particle oscillates between the wells corresponding to $n=+6$ on the right and $n=-6$ on the left of the imaginary axis.}
\end{figure}

\begin{figure}
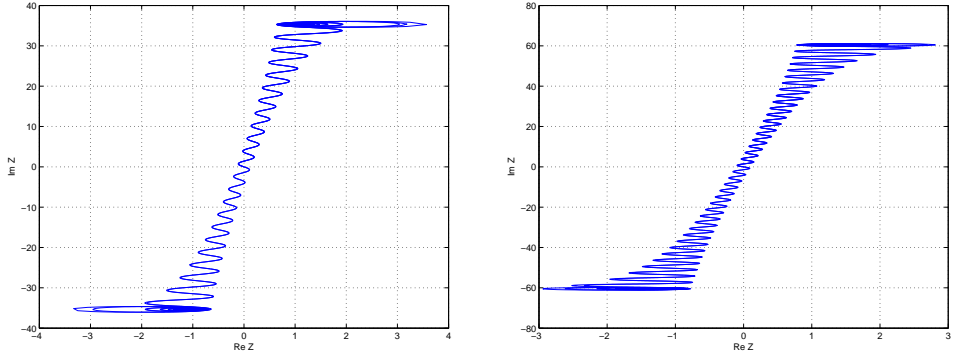

\includegraphics[width=6cm]{Z=1,M=4.epsc} \ \ \ \
\includegraphics[width=6cm]{Z=1,M=5.epsc}
\caption{The trajectory of a particle with energy $E=1+i$ in potential for $\zeta=1$.
{\bf Left}: for ${ M=4}$, the particle oscillates between the wells corresponding to $n=+11$ on the right and $n=-11$ on the left of the imaginary axis.
 {\bf Right}: for ${M=5}$, the particle oscillates between the wells corresponding to  $n=+19$ on the right and $n=-19$ on the left of the imaginary axis.}
\end{figure}

We have performed the numerical study for both odd and even values of $M$ and with
$\zeta$ values below and above the critical value ($\zeta_c$). From the quantum mechanical analysis of this QES problem in Sec. II we  know
that when $M$ is odd and $\zeta\le\zeta_c$ we have unbroken PT phase and QES energy eigenvalues are real. On the other hand PT symmetry breaks
spontaneously  if (i) $M$ is even with any $\zeta $ value or (ii) $M$ is odd with $\zeta > \zeta_c$.
 For broken PT symmetry, one can expect irregular  trajectories with crossing points and disordered behavior of the system. But, as is evident from the trajectories in the figs 7-10 we see in the case of broken PT-symmetry, there is no crossing of trajectories in Non-PT Symmetric potentials.
Further, there are no qualitative differences in the trajectories of the particle in PT-symmetric and Non-PT symmetric potentials classically, unlike the analogous situations in quantum theory.
These results contradict the claims and speculations in Ref \cite{cm1}. However this is not very surprising as these results support the views of the Ref \cite{af}.



\section{Tunneling time }

\setlength{\tabcolsep}{5pt}

\begin{figure}[h]
\includegraphics[width=8cm]{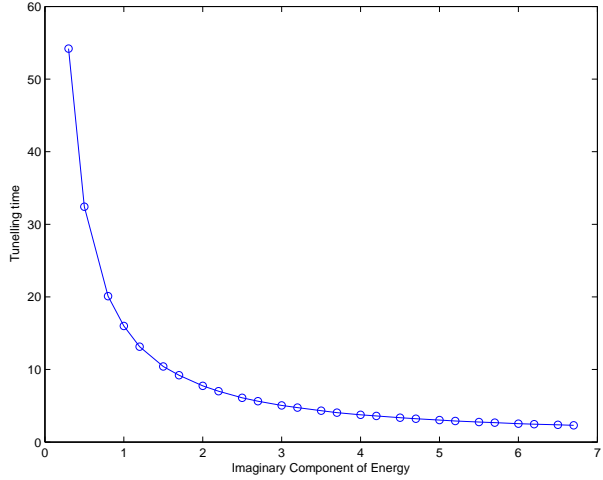}
\caption{\label{fig12}The variation of tunneling time with the imaginary part of energy for a particle initially placed in a potential well corresponding to $n=0$ of PT Symmetric potential ($M=3, \zeta=0.1$) on the right side of the imaginary axis. The real part of energy is fixed as 1 unit.}
\end{figure}

\begin{table}
\caption{The variation of tunneling time across 2 wells with the imaginary component of energy.} \label{table1}
\begin{center}
\begin{tabular}{| c | c |}
\hline
Imaginary Component of Energy Im(E) & Tunneling time  \\
 \hline
	0.3 &				54.19 \\
	0.5 &				32.42\\
	0.8 &				20.1	\\
	1.0 &				15.99\\
	1.2 &				13.14\\
	1.5 &				10.42\\
	1.7 &				9.203\\
	2.0	 &			7.739\\
	2.2	&			7.008\\
 	2.5	&			6.097\\
	2.7	&			5.635\\
	3.0	&			5.054\\
	3.2	&			4.745\\
	3.5	&			4.329\\
	3.7	&			4.058\\
	4.0	&			3.76\\
	4.2	&			3.601\\
	4.5	&			3.355\\
	4.7	&			3.22\\
	5.0	&			3.025\\
	5.2	&			2.902\\
	5.5	&			2.763\\
	5.7	&			2.673\\
	6.0	&			2.541\\
	6.2	&			2.47\\
	6.5	&			2.375\\
	6.7	&			2.313\\
\hline
\end{tabular}
\end{center}
\end{table}

The time taken by the particle to tunnel from one well to another depends on the parameters $M,\zeta$ and the imaginary component of energy($E_2$) of the particle.
We have calculated the tunneling time, that is the time duration between two consecutive crossings of imaginary axis by the particle (in opposite directions)
for fixed $M$ and $\zeta$. The particle spends different amount of time to the left and to the right of the imaginary axis, which is not expected to be the case in classical mechanics. The tunneling time is calculated as the average of the time spent by the particle on both sides of the imaginary axis.
The variation of the averaged tunneling time with $E_2$ is listed in a tabular form [Table 1]. We observe that tunneling time varies inversely with $E_2$ . This  implies quicker tunneling  when $E_2$ is more. The tunneling time is plotted versus $E_2$  (in figure \ref{fig12}).
The nature of this curve satisfies the correspondence principle, since the particle shows more quantum behaviour for
higher $E_2$.
We would also like to point out that no tunneling is seen for the particle with real energy
as it either moves in closed orbits or it escapes to infinity remaining on the same side of the imaginary axis. This tunneling behavior of a classical particle in the complex plane is something which is
generally observed in the domain of quantum physics.

\section{Conclusion}

We have studied the complex classical mechanics of a system whose non-Hermitian PT-invariant version is a QES system.
We treat this model classically on a complex plane  to capture some of the strange behavior of the system. We find that the particle tunnels back and forth between two wells (one on the left side and other on the right of the imaginary axis).
Positions of the wells between which it tunnels back and forth depend on the values of the parameters $M, \zeta$ and the imaginary part of $E$. The distance between the wells in which the particle tunnels, increases with $M$ and it
 decreases when $\zeta$ increases. Particle never tunnels between the wells which are located on the same side of the imaginary axis. The time spent by the particle in the left well is different from the time spent in the right well.
The tunneling time is inversely proportional to the imaginary part of the energy. More $E_2$ implies more tunneling effect and in that situation
tunneling is between nearer wells.
 This can be
 understood by reducing $E_2$ gradually. With very less  $E_2$, the particle will tunnel between wells which are very far.
And for zero imaginary energy, the particle either stays inside the same well with a closed orbit,
or tries to tunnel between 2 wells, which are infinitely separated in the imaginary direction, resulting in an open orbit.
 We do not observe any difference in the overall behavior of the particle dynamics for PT-broken and unbroken situations.

Acknowledgments: One of us (BPM) acknowledges the financial support from Department of Science and Technology (DST), Govt. of India under
SERC project sanction grant No. SR/S2/HEP-0009/2012 and the hospitality of the organizers of
PHHQP11 held at APC, Paris where this work was presented.

\end{document}